\documentclass[prd,twocolumn,showpacs,aps,epsbox,groupedaddress,eqsecnum]{revtex4}
\usepackage{amsmath,amssymb}
\usepackage[dvipdf]{graphicx}

\begin{document}
\draft
\title{Wormhole Shadows}
\author{Takayuki Ohgami}
\email{u006vc@yamaguchi-u.ac.jp}
\author{Nobuyuki Sakai}
\email{nsakai@yamaguchi-u.ac.jp}
\affiliation{Graduate School of Science and Engineering, Yamaguchi University, Yamaguchi 753-8512, Japan}

\begin{abstract}
We propose a new method of detecting Ellis wormholes by use of the images of wormholes surrounded by optically thin dust.
First, we derive steady solutions of dust and more general medium surrounding the wormhole by solving relativistic Euler equations.
We find two types of dust solutions: one is a static solution with arbitrary density profile, and the other is a solution of dust which passes into the wormhole and escapes into the other side with constant velocity.
Next, solving null geodesic equations and radiation transfer equations, we investigate the images of the wormhole surrounded by dust for the above steady solutions.
Because the wormhole spacetime possesses unstable circular orbits of photons, a bright ring appears in the image, just as in Schwarzschild spacetime.
This indicates that the appearance of a bright ring solely confirms neither a black hole nor a wormhole.
However, we find that the intensity contrast between the inside and the outside of the ring are quite different.
Therefore, we could tell the difference between an Ellis wormhole and a black hole with high-resolution very-long-baseline-interferometry observations in the near future.
\end{abstract}
\pacs{04.40.-b, 97.60.Lf}
\maketitle

\section{Introduction}
General relativity and other extended gravitational theories admit a spacetime with nontrivial topology such as a wormhole.
A wormhole is a tunnel-like structure which connects two distant or disconnected regions.
The original wormhole solution, which is called the Einstein-Rosen bridge, was discovered by Einstein and Rosen \cite{ER}.
Because this wormhole is not traversable, it was regarded as nothing but a mathematical product.
Many years later, Ellis \cite{Ellis} obtained a new wormhole solution: a spherically symmetric solution of Einstein equations with a ghost massless scalar field. Morris and Thorne \cite{005} proved that these Ellis wormholes are traversable: instantaneous space movement and time travel could be achieved by passing through the wormhole.
These wormholes have neither singularity nor horizon, and the tidal force is so weak that people can withstand.
If such a wormhole exists, it could become a fascinating tool for voyaging to far galaxies or engaging in time travel.

The stability of traversable wormholes has been studied by several researchers.
Shinkai and Hayward \cite{SH} showed by numerical simulations that Ellis wormholes are unstable.
Gonz\'{a}lez {\it et al.} \cite{GGS} considered more general wormhole solutions with a ghost scalar field and found that they are also unstable.
These results indicate that Ellis wormholes and other traversable wormholes with a ghost scalar field are practically nonexistent.
However, Das and Kar \cite{DK} pointed out that another matter could contribute to supporting the Ellis geometry.
Furthermore, if we adopt modified gravitational theories, a matter like a ghost scalar field, which makes spacetimes unstable, may not be required.
Therefore, a traversable wormhole is still a viable
subject not only in theoretical physics, but also in observational astrophysics.

A possible method for probing wormholes, gravitational lensing of Ellis wormholes has been intensively studied in the literature.
Basic properties of their gravitational lensing were investigated theoretically in Ref.\cite{basic}.
Since Cramer {\it et al.}\ \cite{001} pointed out anomalous features of the light curve of a distant star lensed by a wormhole, observational research to find wormholes by using the microlensing effect has proceeded \cite{002}.
In addition to the light curve, the lensed images \cite{image} and the lensed spectra \cite{spectra} of Ellis wormholes have also been discussed as observable quantities.

As for black holes, another method of probing them by electromagnetic observations is the use of shadows, which are the images of optical or radio sources around a black hole.
Black hole shadows were originally discussed by Bardeen \cite{Bardeen} and have recently attracted much attention \cite{Takahashi}.
Black hole shadows have not only been studied theoretically but have also applied to probing black holes by very-long-baseline-interferometry (VLBI) observations \cite{VLBI}.
Therefore, we expect that shadows could also be used to probe wormholes by VLBI observations.
The shadows of a rotating wormhole were studied by Nedkova, Tinchev, and Yazadjiev \cite{The Shadow of a Rotating Traversable Wormhole}. They calculated the outline of the shadow in the presence of an extended source behind the wormhole.

In this paper we consider the images of wormholes surrounded by optically thin dust.
Recently it was shown that Ellis wormholes possess unstable circular orbits of photons \cite{orbit}, as we discuss in Sec. IV.
Because we expect a bright ring to appear in the image, just as in Schwarzschild spacetime, it is important to explore whether it is possible to identify wormholes by observing shadows.

This paper is organized as follows. In Sec. II, we introduce the Ellis wormhole and discuss its spacetime structure briefly.
In Sec. III, to set up dust models used in our shadow analysis, we derive steady solutions of dust and a more general medium surrounding the wormhole by solving relativistic Euler equations.
In Sec. IV, we derive null geodesic equations and discuss photon trajectories around the wormhole.
In Sec. V, we investigate---by solving the radiative transfer equation as well as the null geodesic equations---the images of the wormhole surrounded by dust for the models obtained in Sec. II.
Section VI is devoted to concluding remarks.
%
%
%
%
%
\section{Ellis wormhole spacetime}
\begin{figure}[h]
\centering
\includegraphics[width=.7\hsize]{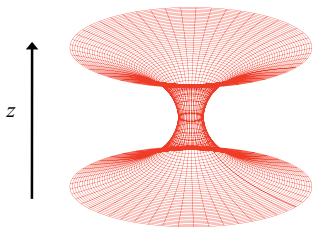}
\caption{Embedding diagram of the Ellis wormhole: plot of $z$ vs. $r^*$ in (\ref{eq:embed equation in 3D space of 2D wormhole}) with the $\varphi$ direction.}
\label{fig:2D wormhole image}
\end{figure}

The Ellis wormhole spacetime is one of the traversable ones, and it is expressed by the line element
\begin{align}
	ds^2=-dt^2+dr^2+(r^2+a^2)(d\theta^2+\sin^2\theta\,d\varphi^2),
	\label{eq:Ellis wormhole line element}
\end{align}
where  $a$ is the throat radius of the wormhole.
This line element indicates that there is no singularity in this spacetime.
To understand the geometry of the spacetime intuitively, we draw an embedding diagram as follows.
Introducing a new radial coordinate as ${r^*}^2=r^2+a^2$, we express the two-dimensional sphere of $t,~\theta =\text{const}$ as
\begin{align}
	ds^2=\frac{d{r^*}^2}{1-a^2/{r^*}^2}+{r^*}^2d\varphi^2.
	\label{eq:Ellis wormhole line element(2)}
\end{align}
Suppose a three-dimensional Euclidean space which is described by cylindrical coordinates,
\begin{equation}
	ds^2=d{r^*}^2+{r^*}^2d\varphi^2+dz^2.
	\label{Euclid}
\end{equation}
Comparing (\ref{Euclid}) and (\ref{eq:Ellis wormhole line element(2)}), we obtain the relation
\begin{align}
	z=\pm a\cdot\text{arccosh}\frac{{r^*}}{a}.
	\label{eq:embed equation in 3D space of 2D wormhole}
\end{align}
Figure \ref{fig:2D wormhole image} is a graph of (\ref{eq:embed equation in 3D space of 2D wormhole}) with the $\varphi$ direction, which indicates 
the visual image of the spacetime structure of the Ellis wormhole.
We see that two separated spaces are connected by the throat like a tunnel.

\section{Motion of interstellar medium}
In this section, we examine the general relativistic motion of the interstellar medium around the Ellis wormhole under the assumption that its self-gravity is negligible.
The equations of motion of the perfect fluid in Schwarzschild spacetime and their solutions were presented in a textbook by Shapiro and Teukolsky \cite{BLACK HOLES WHITE DWARFS AND NEUTRON STARS THE PHYSICS OF COMPACT OBJECTS}.
Here we apply their method to the Ellis wormhole.

\subsection{Thermodynamics}
Suppose a local Lorentz frame comoving with fluid particles.
Let $n, \rho$ and $P$ be, respectively, the number density, the total energy density, and the pressure of the fluid particles, which are measured in the reference frame.
Then the first law of thermodynamics is written as
\begin{align}
	dQ =d\left(\frac{\rho}{n}\right)+Pd\left( \frac{1}{n} \right),
	\label{eq:First law of thermodynamics}
\end{align}
where $dQ$ is the heat gained per particle. 
$\rho/n$ and $1/n$ represent, respectively, the energy and the volume par particle.
If the process is quasistatic (i.e., in thermal equilibrium at all times) and adiabatic, then $dQ=Tds=0$,  where $s$ is the entropy per particle and $T$ is the temperature.
It follows that
\begin{align}
	0=d\left(\frac{\rho}{n}\right)+Pd\left( \frac{1}{n} \right).
	\label{eq:total differential of the internal energy(2)}
\end{align}
This equation displays energy conservation and is rewritten in the following, simpler form:
\begin{align}
	\frac{d\rho}{dn}=\frac{\rho+P}{n}.
	\label{eq:mass-energy conservation(2)}
\end{align}

\subsection{Fluid dynamics}
Assuming that the particle number is conserved, we have the general relativistic continuity equation,
\begin{align}
	\left( nu^\mu\right)_{;\mu}=0,
	\label{eq:continuity equation}
\end{align}
where $u^\mu$ is the four-velocity of the particle fluid.
Here the semicolon denotes a covariant derivative: $A^{\alpha}_{;\beta}=\partial_{\beta}A^{\alpha}+\Gamma^{\alpha}_{\beta \gamma}A^{\gamma}$.
We also assume that the particle fluid is a perfect fluid whose energy-momentum tensor is written as
\begin{align}
	T^{\mu\nu}=(\rho +P)u^{\mu}u^{\nu}+Pg^{\mu\nu}.
	\label{eq:energy-momentum tensor}
\end{align}
Applying the conservation law of energy-momentum,
\begin{align}
	T^{\nu}_{\mu;\nu}=0,
	\label{eq:Conservation law of energy-momentum}
\end{align}
to (\ref{eq:energy-momentum tensor}) with (\ref{eq:mass-energy conservation(2)}), we obtain the general relativistic Euler equation,
\begin{align}
	(\rho+P)u_{\mu;\nu}u^{\nu}=-P_{\mu}-u_{\mu}P_{,\nu}u^{\nu}.
	\label{eq:relativistic Euler equation}
\end{align}
We have two basic equations (\ref{eq:continuity equation}) and (\ref{eq:relativistic Euler equation}), for the general perfect fluid.

\subsection{Spherical steady-state solutions}
Here we suppose spherically symmetric flow without angular momentum.
This assumption is just for the simplicity of obtaining analytic solutions for dust flow.
Then the four-velocity of the fluid takes the form
\begin{align}
	u^{\alpha}=\frac{dx^{\alpha}}{d\tau}=(u^t,\,u^r,\,0,\,0),
	\label{eq:4D flow velocity??component}
\end{align}
where $\tau$ is the proper time of the fluid.
For the metric (\ref{eq:Ellis wormhole line element}), the continuity equation (\ref{eq:continuity equation}) and the relativistic Euler equations (\ref{eq:relativistic Euler equation}) become 
\begin{align}
	\frac{\partial n}{\partial t}u^t+\frac{\partial n}{\partial r}u^r+n\left(\frac{\partial u^t}{\partial t}+\frac{\partial u^r}{\partial r}+\frac{2r}{r^2+a^2}u^r \right)=0.
	\label{eq:equation of continuing (2)}
\end{align}
\vspace{-1.5em}
\begin{multline}
	(\rho+P)\left( \frac{\partial u^r}{\partial t}u^t+\frac{\partial u^r}{\partial r}u^r \right) \\
	+\frac{\partial P}{\partial r}(1+{u^r}^2)+\frac{\partial P}{\partial t}u^ru^t=0.
	\label{eq:relativistic Euler equation (2)}
\end{multline}
We also assume that the flow is stationary, that is, $\partial / \partial t=0$.
By use of the relation, $-1=u_\mu u^\mu=-(u^t)^2+(u^r)^2$, we rewrite (\ref{eq:equation of continuing (2)}) and (\ref{eq:relativistic Euler equation (2)}) as
\begin{gather}
	\frac{n'}{n}+\frac{u'}{u}+\frac{2r}{r^2+a^2}=0
	\label{eq:equation of continuing (3)}\\
	uu'=-\frac{1}{\rho+P}\frac{dP}{dr}(1+u^2),
	\label{eq:relativistic Euler equation (3)}
\end{gather}
where $'\equiv d / dr,\,u\equiv|u^r|$.

We integrate the differential equations (\ref{eq:equation of continuing (3)}) and (\ref{eq:relativistic Euler equation (3)}) and obtain the form of conservation equations
\begin{gather}
	\dot{M}=4\pi mnu(r^2+a^2)=\text{const},
	\label{eq:solution (1)}\\
	\left( \frac{\rho+P}{n} \right)^2(1+u^2)=\left( \frac{\rho_{\infty}+P_{\infty}}{n_{\infty}} \right)^2(1+{u_{\infty}}^2),
	\label{eq:solution(2)}
\end{gather}
where the subscript $\infty$ denotes quantities at infinity and $\dot{M}$ is the rest-mass accretion rate.
Because Eqs. (\ref{eq:solution (1)}) and (\ref{eq:solution(2)}) include four unknown variables---the particle number density $n$, the energy density $\rho$, the pressure $P$, and the velocity $u$---we need additional conditions to obtain explicit solutions.

\begin{figure}[h]
	\centering
	\includegraphics[width=1.0\hsize]{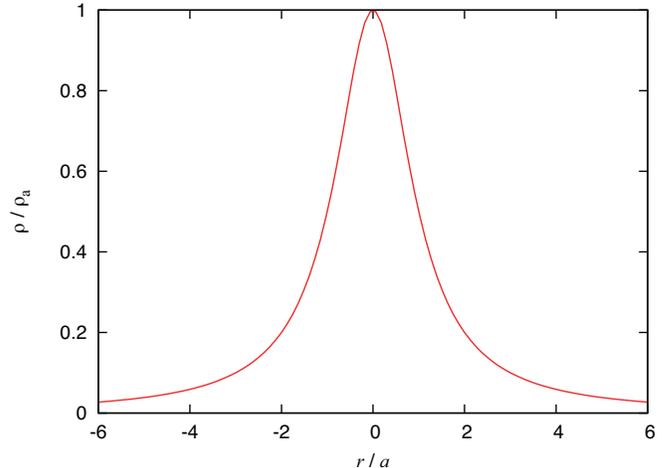}
	\caption{Density distribution of dust for $u\ne0$, which is in accordance with (\ref{eq:dust solution(2)}).  $\rho_{\text{a}}$ is dust density at the throat, $r=0$. The region $r<0$ corresponds to the opposite side of the wormhole.}
	\label{fig:density ratio}
\end{figure}

Hereafter, we assume that the Ellis wormhole is surrounded by dust, which is characterized by 
\begin{align}
	\rho=nm\,,\quad P=0.
	\label{eq:dust condition}
\end{align}
With this condition, we obtain the steady flow solutions
\begin{gather}
	u=u_{\infty}=\text{const.},
	\label{eq:dust solution (1)}\\
	\dot{M}=4\pi \rho u_{\infty}(r^2+a^2)=\text{const}.
	\label{eq:dust solution(2)}
\end{gather}
We find two types of solutions, as follows:
\begin{itemize}
\item $u=u_{\infty}=0,~\rho=$ arbitrary function of $r$.
\item $u=u_{\infty}\ne0,~\rho\propto(r^2+a^2)^{-1}$.
\end{itemize}
The second type implies the solution that dust passes into the wormhole and escapes into the other side with constant velocity, and its reverse.
For this case we show the profile of $\rho(r)$ in Fig. \ref{fig:density ratio}.
Note that the region $r<0$ corresponds to the opposite side of the wormhole.

\section{Photon trajectories}
In preparation for the investigation of wormhole shadows in the next section, we derive the null geodesic equations and discuss qualitative properties of photon trajectories.

\subsection{Null geodesic equations and effective potential}
\begin{figure}[h]
	\centering
	\includegraphics[width=1.0\hsize]{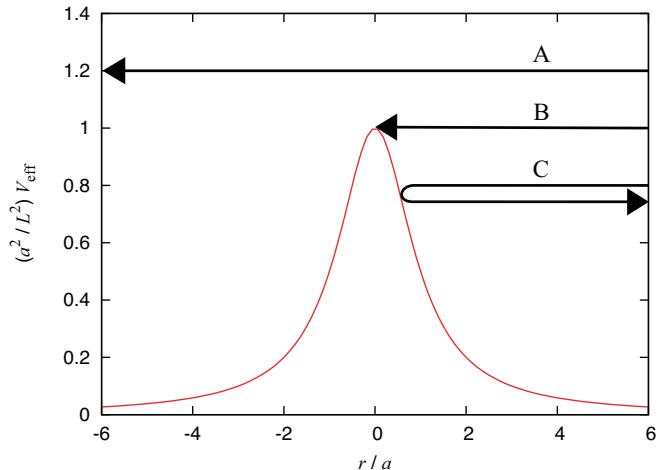}
	\caption{Effective potential of the Ellis wormhole.  The maximum point at $r=0$ corresponds to the unstable circular orbits of photons.}
	\label{fig:Effective potential of wormhole}
\end{figure}

Null geodesic equations are generally written as
\begin{align}
	\frac{dk^{\mu}}{d\lambda}+\Gamma^{\mu}_{\nu\rho}k^{\nu}k^{\rho}=0,\quad \text{with}\quad k_{\mu}k^{\mu}=0,
	\label{eq:geodesic equations}
\end{align}
where $\lambda$ and $k^{\mu}\equiv dx^{\mu}/d\lambda$ are the affine parameter and the null vector, respectively.
The geodesics in the $\theta=\pi/2$ plane are given by
\begin{gather}\label{geo1}
{d\over d\lambda}k^t=0,~~~
{d\over d\lambda}\{(r^2+a^2)k^\varphi\}=0,\\
{d\over d\lambda}k^r-r(k^\varphi)^2=0,
\label{geo2}\\
-(k^t)^2+(k^r)^2+(r^2+a^2)(k^\varphi)^2=0.
\label{geo3}\end{gather}
Because Eq.(\ref{geo2}) is also derived by (\ref{geo1}) and (\ref{geo3}), we do not have to solve it.
By integrating (\ref{geo1}), we obtain two conserved quantities,
\begin{align}
	E=k^t,\quad L=(r^2+a^2)k^{\varphi}
	\label{eq:Conserved quantity}
\end{align}
Then (\ref{geo3}) becomes
\begin{align}
	{k^r}^2+V_{\text{eff}}(r)=E^2,\quad V_{\text{eff}}(r)\equiv\frac{L^2}{r^2+a^2}.
	\label{eq:effective potential of wormhole}
\end{align}
We show the effective potential  $V_{\text{eff}}$ in Fig.\ \ref{fig:Effective potential of wormhole}, which will be discussed in the next subsection.

It follows from (\ref{eq:Conserved quantity}) and (\ref{eq:effective potential of wormhole}) that
\begin{equation}
{dr\over d\varphi}={k^r\over k^\varphi}
=\pm{r^2+a^2\over L}\sqrt{E^2-V_{\text{eff}}},
\label{drdphi}\end{equation}
which gives spatial trajectories of photons, which is discussed in the next subsection.
For use in Sec.\ V, we derive the relation from (\ref{drdphi}),
\begin{equation}
\lim_{r\rightarrow\infty}{1\over r^2}{dr\over d\varphi}=\pm\frac EL.
\label{eq:E/L}
\end{equation}

\subsection{Photon trajectories}
Photon trajectories in the Ellis geometry were studied in Ref. \cite{orbit}.
Here we reanalyze them and discuss their characteristics.

Using the effective potential shown in Fig.\ \ref{fig:Effective potential of wormhole}, we can discuss qualitative properties of photon trajectories in the wormhole spacetime. 
The trajectories are classified into three types: A, B, and C.
The maximum at the throat $r=0$ corresponds to the unstable circular orbit of photons.
Type C represents the photon which passes by the throat.
Type A represents the photon which passes into the throat and escapes into the other side.
Type B represents the photon which approaches and winds many times in the vicinity of the unstable circular orbit.

\begin{figure}[htbp]
  \begin{center}
    \begin{tabular}{c}
      \begin{minipage}{.95\hsize}
        \begin{center}
          \includegraphics[scale=1.05]{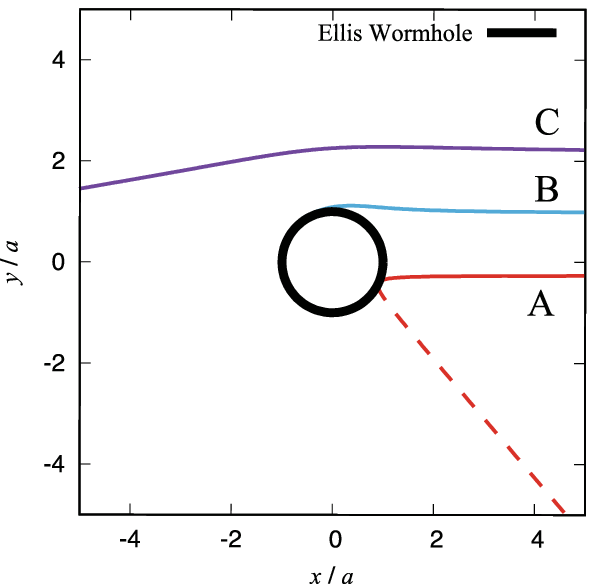}
          \hspace{3cm} (a)
    	  \vspace{0.3cm}
        \end{center}
      \end{minipage}\\
      \begin{minipage}{.95\hsize}
        \begin{center}
          \includegraphics[scale=1.09]{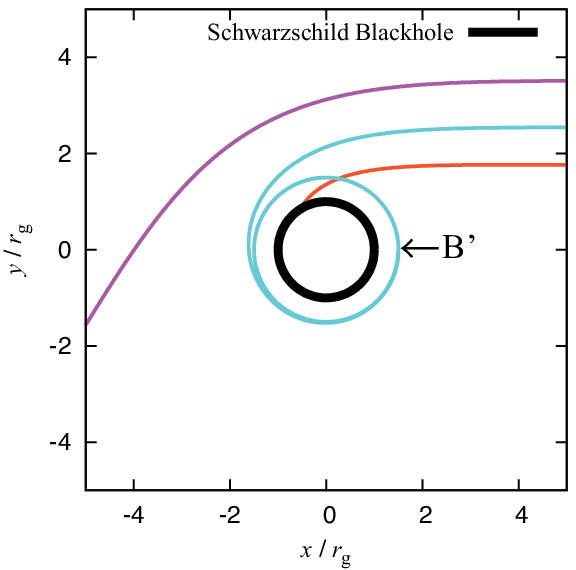}
          \hspace{3cm} (b)
        \end{center}
      \end{minipage}
    \end{tabular}
    \caption{Photon trajectories around (a) the Ellis wormhole and (b) the Schwarzschild black hole. 
    The coordinates $(x,~y)$ are defined by (\ref{xy}) on the $\theta=\pi/2$ plane.
   These trajectories end up at the observer at $r=300a,~\varphi=0$.
    In (a) labels A, B, and C correspond to those in the effective potential in Fig.\ 3;
    the dashed line of A represents the trajectory on the other side of the wormhole ($r<0$).
    In (b) $r_g$ denotes the Schwarzschild radius;
    the trajectory B' represents the photon which approaches and winds many times in the vicinity of the unstable circular orbit.
}
    \label{fig:orbit_WH/BH}
  \end{center}
\end{figure}

Figure \ref{fig:orbit_WH/BH}(a) shows photon trajectories around the wormhole.
On the $\theta=\pi/2$ plane we define the rectangle coordinates $(x,~y)$  as
\begin{equation}\label{xy}
x=r^*\cos\varphi,~~~y=r^*\sin\varphi.
\end{equation}
These trajectories end up at the observer at $r=300a,~\varphi=0$.
Labels A, B, and C correspond to those on the effective potential in Fig.\ 3.
The dashed line of A represents the trajectory in the other side of the wormhole ($r<0$).

For reference, we also show trajectories around a Schwarzschild black hole in Fig. \ref{fig:orbit_WH/BH}(b).
The trajectory B$'$ in Fig. 4(b) represents the photon which approaches and winds many times in the vicinity of the unstable circular orbit, which is analogous to the trajectory B in Fig. 4(a). Accordingly, in both spacetimes one could observe brightening there when gas falls into the wormhole or the blackhole.
This similarity is important, but we are also interested in the observational difference between the two.

\section{Wormhole shadows}
In this section, we investigate the optical images of the wormhole surrounded by dust, using the solutions obtained in Sec.\ III.

\subsection{Apparent position of optical sources}
\begin{figure}[h]
	\centering
	\includegraphics[width=1.0\hsize]{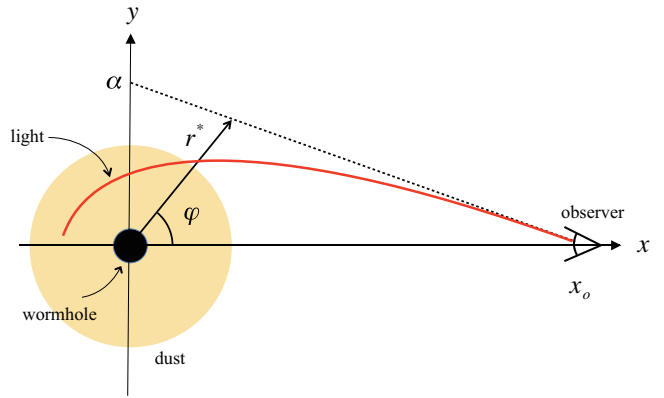}
	\caption{Light orbit in the $\theta=\pi/2$ plane. We adopt the rectangle coordinates $(x,~y)$ defined by (\ref{xy}).
	The center of the wormhole is located at the origin and the observer at $(x_o,0)~(\varphi=0)$. 
	We denote the intersection of the $y$ axis with the tangent to the ray at the observer by $y=\alpha$.
        Therefore, we can regard $\alpha$ as the apparent length from the center.
}
	\label{fig:Diagram of light orbit in Ellis wormhole}
\end{figure}

We consider the rectangle coordinates $(x,~y)$ defined by (\ref{xy}) on the $\theta=\pi/2$ plane, where the center of the wormhole
is located at the origin and the observer at $(x_o,0)~(\varphi=0)$, as shown in Fig.\ \ref{fig:Diagram of light orbit in Ellis wormhole}.
We denote the intersection of the $y$ axis with the tangent to the ray at the observer by $y=\alpha$.
Therefore, we can regard $\alpha$ as the apparent length from the center.

The equation of this tangent line is
\begin{equation}\label{tangent}
{x\over x_o}+{y\over\alpha}=1,~~\text{i.e.},~~
{r^*\over x_o}\cos\varphi+{r^*\over\alpha}\sin\varphi=1.
\end{equation}
Differentiating (\ref{tangent}) and taking the limit of $r^*\rightarrow x_o,~\varphi\rightarrow0$ , we obtain
\begin{equation}
\alpha=-x_o^2{d\varphi\over dr^*}.
\end{equation}
Furthermore, taking the limit of $x_o\rightarrow\infty$ and using (\ref{eq:E/L}), we find
\begin{equation}\label{LE}
\alpha\rightarrow\frac LE.
\end{equation}
As Fig.\ 5 shows, $L/E$ represents the apparent positions of optical sources; therefore, it is interpreted as an impact parameter.

We numerically integrate the null geodesic equations from the observer to the sources.
Once we specify the initial point and the orbit plane, the remaining parameter of the geodesic equations is the impact parameter $L/E$ only.

\subsection{Radiation intensity}
To calculate the observed intensity of radiation emitted from optically thin gas, we solve the general relativistic radiative transfer equation, which is generally expressed as  \cite{FOUNDATIONS OF RADIATION HYDRODYNAMICS},
\begin{align}
	\frac{d\, \mathfrak{J}}{d\lambda}=\frac{\eta(\nu)}{\nu^2}-\nu\chi(\nu)\mathfrak{J},\qquad \mathfrak{J}\equiv\frac{I(\nu)}{\nu^3}
	\label{eq:radiative transport equation}
\end{align}
where $\nu$ is the photon frequency, $I(\nu)$ is the specific intensity, $\mathfrak{J}$ is the invariant intensity, $\eta(\nu)$ is the emission coefficient and $\chi(\nu)$ is the absorption coefficient.
Because Eq.(\ref{eq:radiative transport equation}) is the differential equation along null geodesics, we should solve the null geodesic equations simultaneously.

Here we make the following assumptions, for simplicity:
\begin{itemize}
\item The dust does not absorb radiation, i.e., $\chi(\nu)=0$.
\item $\eta(\nu)$ is proportional to the dust density which is measured along the null geodesics, i.e., $\eta(\nu) d\lambda \propto\rho u_\mu dx^{\mu}$.
\end{itemize}
Introducing a positive factor $H(\nu)$, which is proportional to the spectrum of the dust sources, we express $\eta(\nu)$ as
\begin{align}
	\eta(\nu)d\lambda =-H(\nu)\rho u_{\mu}dx^{\mu}.
	\label{eq:radiation term}
\end{align}
With these assumptions we can integrate (\ref{eq:radiative transport equation}) as
\begin{align}
	\mathfrak{J}=-\int\frac{H(\nu)}{\nu^2}\rho u_{\mu}dx^{\mu}.
	\label{eq:integration of transfer equation}
\end{align}
 The integration in (\ref{eq:integration of transfer equation}) should be performed alongside the null geodesics.
The frequency measured by observers comoving with the dust particles is given by
\begin{equation}
\nu=-u_\mu k^\mu.
\end{equation}

Generally we should fix the spectrum of the dust sources, i.e., $H(\nu)$.
Here, for simplicity, we assume a flat spectrum, $H(\nu)=$const.

\subsection{Numerical analysis}
We numerically calculate the intensity distribution as follows:
\def\theenumi{(\roman{enumi})}
\begin{enumerate}
	\item Put the observer at $x_o=300a$.
	\item For a given value of $\alpha=L/E$, we solve the null geodesic equations from the observer.  We can choose a value of the initial (observed) frequency $\nu_o$ arbitrarily because the ratio of $\nu_o$ to the emitted frequency $\nu_e$ does not depend on $\nu_o$.
	\item With the values of $\nu$ at each point, which is determined by the null geodesic equations, we integrate (\ref{eq:integration of transfer equation}) to obtain the intensity $I$. We adopt the fourth-order Runge-Kutta method for all integrations.
	\item We continue the integrations until $r=300a$ again, where the gas density is sufficiently small.
	\item Iterate (ii) $\sim$ (iv) by changing the value of $\alpha$.
\end{enumerate}

\begin{figure}
\begin{center}
          \includegraphics[scale=.9]{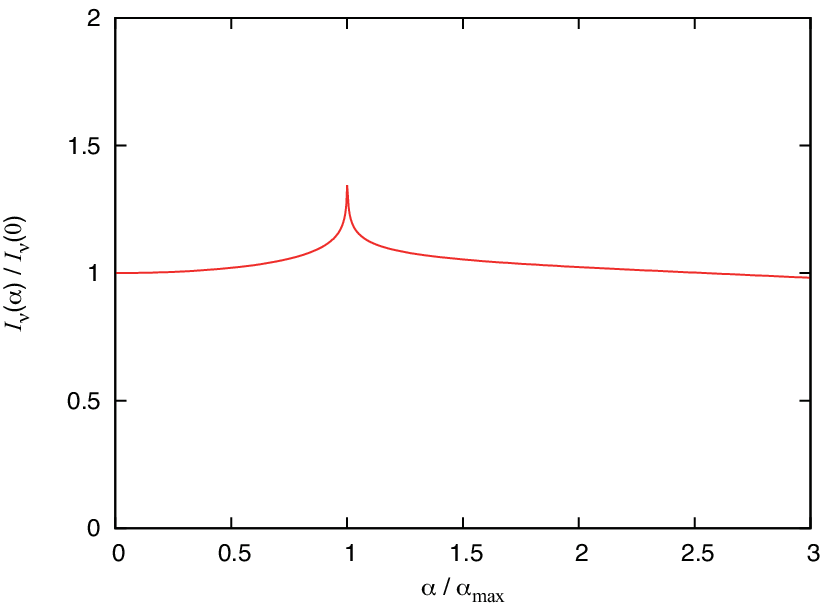}\\
          (a)
          \\ \vspace{.3cm}
          \includegraphics[scale=.3]{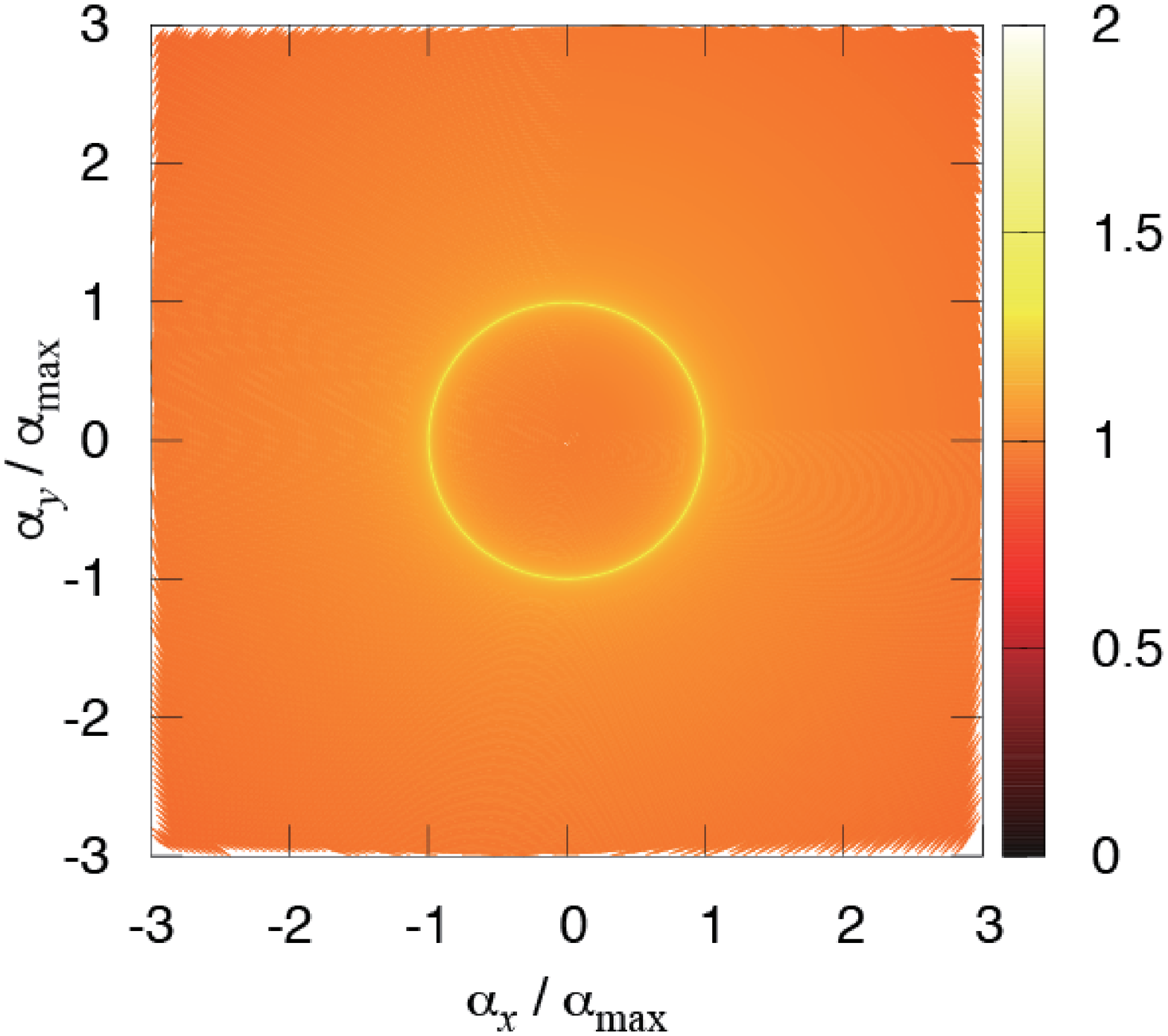}\\
          (b)\\
          \vspace{-.5cm}
 \end{center}
    \caption{Radiation intensity for case 1:  static top-hat distribution. (a) is the graph of $I_\nu(\alpha)/I_\nu(0)$, which
    does not depend on $\nu$ for the flat spectrum, $H(\nu)=$const. $\alpha_{{\rm max}}$ denotes the value of $\alpha$ where $I_\nu(\alpha)$ is the maximum; it corresponds to the photon which winds around the throat infinite times.
    (b) is the optical image, which is depicted based on (a).
    The color shows intensity and is irrelevant to the frequency. }
    \label{fig:Numerical calculation result of Case 1}
\end{figure}
\begin{figure}
\begin{center}
          \includegraphics[scale=.9]{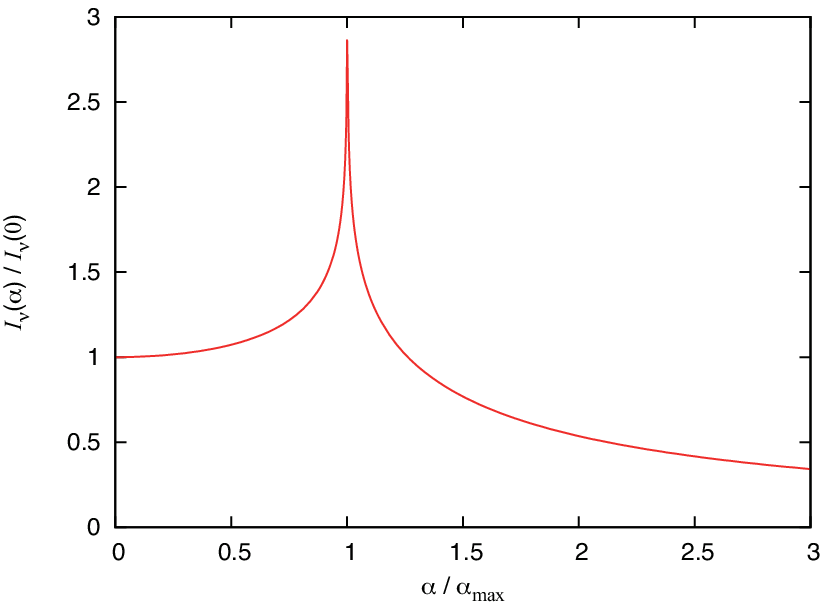}\\
          (a)
          \\ \vspace{.3cm}
          \includegraphics[scale=0.3]{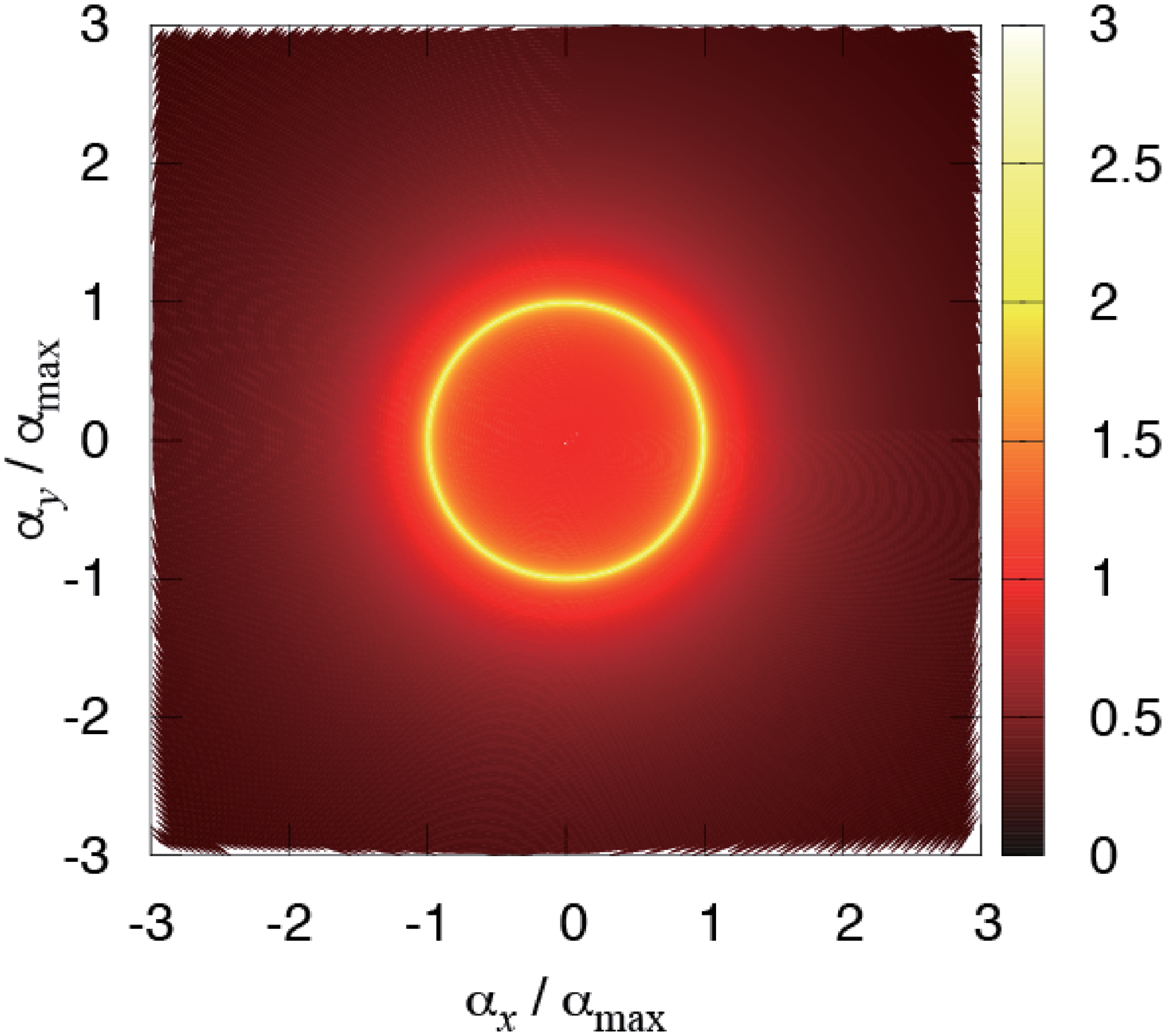}\\
          (b)\\
          \vspace{-.5cm}
 \end{center}
    \caption{Radiation intensity for case 2: nonstatic dust that flows into the wormhole and escapes into the other side. As in Fig. 6, (a) shows intensity as a function of $\alpha$ and (b) displays the optical image.}
    \label{fig:Numerical calculation result of Case 2}
\end{figure}

As for the dust distribution,  we consider two cases.
\begin{itemize}
\item Case 1: static top-hat distribution.
	$$u=0,~~~\rho=
	\begin{cases}
	{\rm const}>0 & (-10<r/a<10)\\
	0 & ({\rm otherwise})
	\end{cases}$$
\item Case 2: nonstatic dust that flows into the wormhole and escapes into the other side.
	$$u=10^{-7},~~~\rho\propto\{(r/a)^2+1)\}^{-1}.$$
\end{itemize}

Figures \ref{fig:Numerical calculation result of Case 1} and \ref{fig:Numerical calculation result of Case 2} show the numerical results for radiation intensity for cases 1 and 2, respectively.
Figures 6(a) and 7(a) are the graphs of $I_\nu(\alpha)/I_\nu(0)$. 
In general, they depend on $\nu$; however, they do not for the flat spectrum, $H(\nu)=$const. 
$\alpha_{{\rm max}}$ denotes the value of $\alpha$ where $I_\nu(\alpha)$ is the maximum; it corresponds to the photon which winds around the throat infinite times.
Figures 6(b) and 7(b) display the optical images. In both cases a bright ring appears, in accord with the peak in Figs. 6(a) and 7(a).

We also investigate case 2 for various values of the fluid velocity $u$. As long as the fluid is nonrelativistic (i.e., $|u|\ll1$), the result does not change essentially even if the flow is in the reverse direction (i.e., $u<0$). We may therefore conclude that the profile and appearance in Fig.\ \ref{fig:Numerical calculation result of Case 2} is the general feature for nonstatic dust that flows into or out of the wormhole.

As we discussed in Sec. IV.B, the appearance of a bright ring is a characteristic common to black holes.
This indicates that the appearance of a bright ring solely confirms neither a black hole nor a wormhole.
Therefore, we closely reinvestigate the image of the Schwarzschild black hole surrounded by optically thin dust.
The stationary solution of dust flow in the Schwarzschild spacetime is expressed as \cite{BLACK HOLES WHITE DWARFS AND NEUTRON STARS THE PHYSICS OF COMPACT OBJECTS}
\begin{gather}
	u\propto r^{-\frac12},\quad 
	\rho\propto r^{-\frac{3}{2}}.
	\label{eq:dust solution in Schwarzschild spacetime}
\end{gather}

\begin{figure}
\begin{center}
	\vspace{0.3cm}
          \includegraphics[scale=.9]{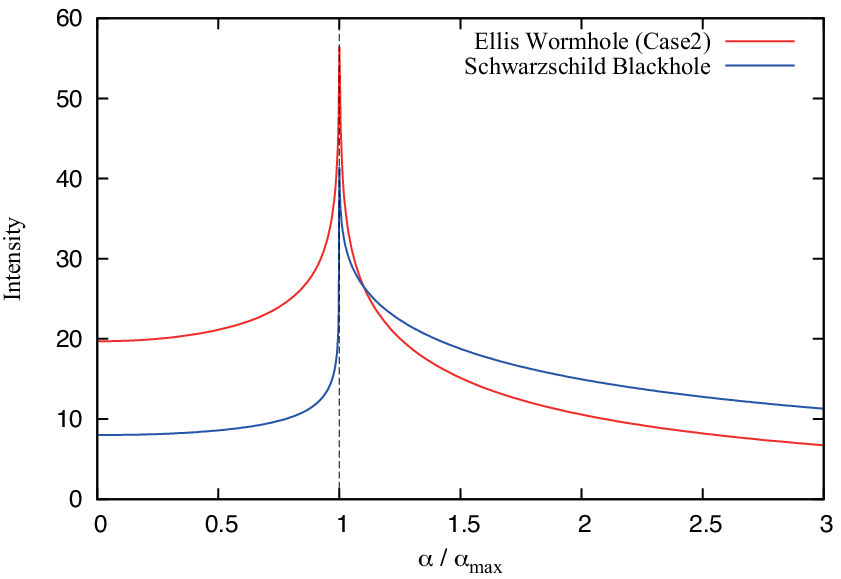}\\
          (a)
          \\ \vspace{.3cm}
          \includegraphics[scale=0.3]{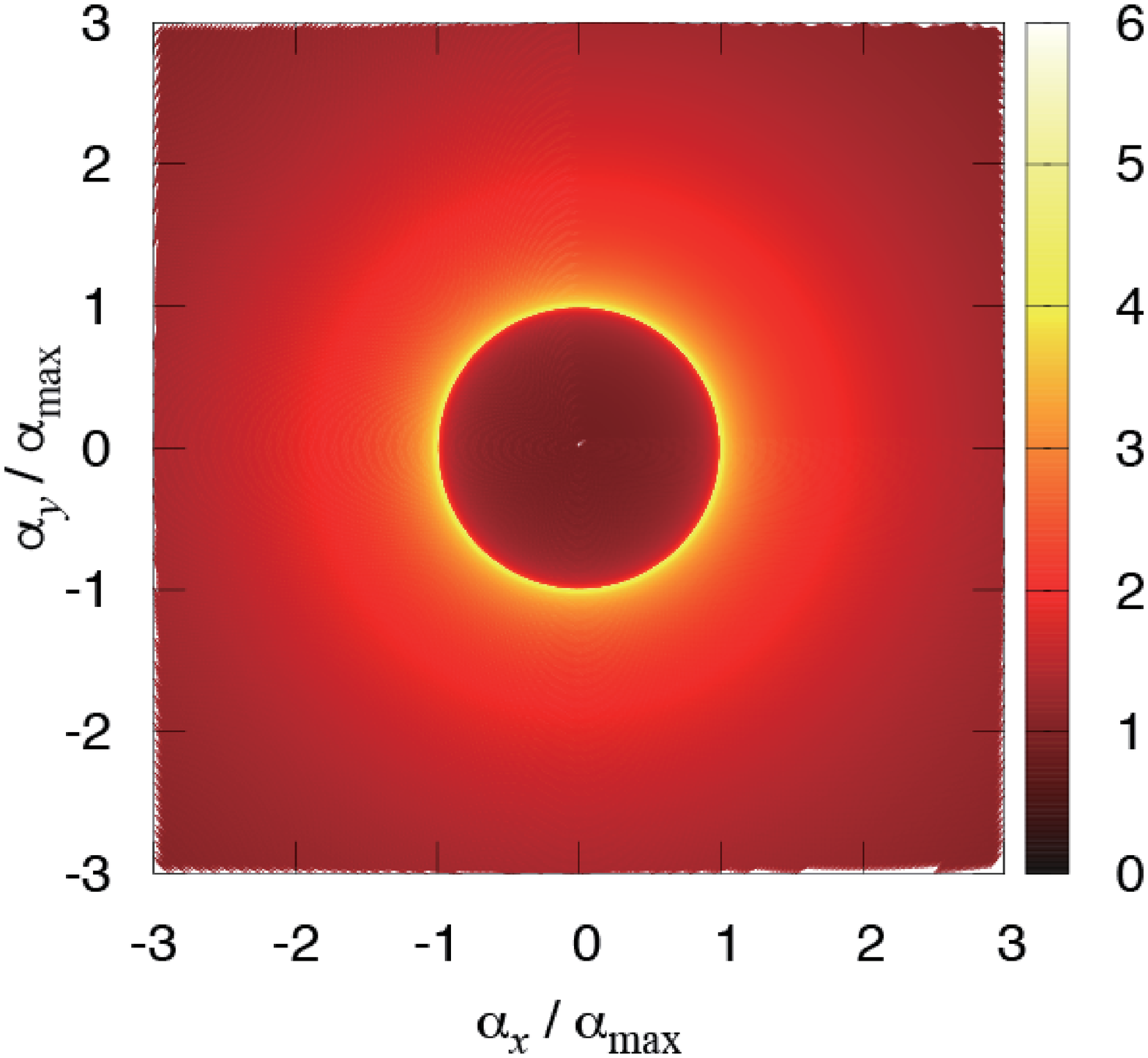}\\
          (b)\\
          \vspace{-.5cm}
 \end{center}
 \caption{Radiation intensity for dust flowing into the Schwarzschild black hole. In (a), for comparison, we plot $I_{\nu}$ for the wormhole (case 2) as well as that for the black hole. They are rescaled so that the difference in shape between them is clear. (b) displays the optical image for the black hole.}
	\label{fig:Comparison of wormhole and blackhole}
 \end{figure}

Figure \ref{fig:Comparison of wormhole and blackhole} shows the numerical results of radiation intensity for the above dust model.
For comparison, we also plot the result for the wormhole (case 2) in Fig. 8(a).
While the two profiles at the peak [and the corresponding bright rings in Fiig. 8(b)] are analogous, the intensity contrast between the inside and the outside of the ring are different.
In the case of the wormhole, because photons are also emitted from dust beyond the throat, the inside looks brighter than the outside.
This contrast could distinguish a wormhole from a black hole with high-resolution radio observations.

%
%
%
%
\section{Concluding remarks}

We have proposed a new method for detecting Ellis wormholes by use of the images of wormholes surrounded by optically thin dust.
First, we derived steady solutions of dust and a more general medium surrounding the wormhole by solving relativistic Euler equations.
We found two types of dust solutions: one is a static solution with an arbitrary density profile, and the other is a solution for dust which passes into the wormhole and escapes into the other side with constant velocity.
Next, solving null geodesic equations and radiation transfer equations, we investigated the images of the wormhole surrounded by dust for the above steady solutions.
Because the wormhole spacetime possesses unstable circular orbits of photons, a bright ring appears in the image, just as in Schwarzschild spacetime.

We have assumed that the dust distribution is spherically symmetric and steady state.
This is just for simplicity to obtain analytic solutions of dust flow;
if we consider more realistic situations, we should obtain numerical solutions of rotating dust flow.
On the other hand, the effect of the angular velocity on radiation flux is negligible as long as the velocity is nonrelativistic.

Our results indicate that the appearance of a bright ring solely confirms neither a black hole nor a wormhole.
This could be a serious problem for identifying black holes by optical/radio observations.
However, we found that the intensity contrast between the inside and the outside of the ring are quite different.
Therefore, we could be able to tell the difference between an Ellis wormhole and a black hole with high-resolution very-long-baseline-interferometry observations in the near future.

%
%
%
%
%
\acknowledgements
We thank F. Abe, K. Fujisawa, T. Harada, H. Saida and K. Shiraishi for useful discussions.

%
%
%
%
%

\end{document}